\documentclass[5p]{elsarticle}
\usepackage[utf8]{inputenc}

\usepackage{amsmath}
\usepackage{amssymb}
\usepackage{commath}
\usepackage{siunitx}

\usepackage{graphicx}

\usepackage{tikz-cd}

\usepackage{bm}

\usepackage{url}

\usepackage{blindtext}

\newcommand{\tabref}[1]{Table~\ref{#1}}
\renewcommand{\eqref}[1]{Eq.~(\ref{#1})}

\newcommand*{\refcite}{}
\DeclareRobustCommand*{\refcite}[1]{%
  \begingroup
    \romannumeral-`\x 
    \setcitestyle{numbers}%
    Ref.~\cite{#1}%
  \endgroup
}

\newcommand{\defeq}{\stackrel{\text{def}}{=}}

\newcommand*{\variable}{\makebox[1ex]{\textbf{$\cdot$}}}

\providecommand*{\eu}{\ensuremath{\mathrm{e}}}

\newcommand{\ie}{i.e.,} 

\DeclareMathOperator{\expone}{E_1}

\begin{document}

\begin{frontmatter}

\title{The differential build-up factor}

\author[usal]{G.~Hern\'andez\corref{cor}}
\ead{guillehg@usal.es}
\author[cea]{A.~Canas-Junquera}
\author[usal]{F.~Fern\'andez}

\cortext[cor]{Corresponding author}

\address[usal]{IUFFyM, Universidad de Salamanca, Salamanca, Spain}
\address[cea]{CEA/Cadarache, Saint-Paul-lès-Durance, Cedex, France}



\sisetup{range-units=single}


\begin{abstract}
The build-factor is a magnitude which allows to correct the photon exponential attenuation model to obtain the real value of a certain dosimetric magnitude, like air exposure. Its main weaknesses are the dependences on the response function of such a magnitude, the geometry, and the source directionality. A formalism to deal with the first is presented in this work, which leads to the definition of the differential build-up factor. Analog Monte Carlo calculations were used to calculate this factor, as well as describing its uncertainty, in the 30 keV to 10 MeV energy range for the most typical materials up to 10 mean-free-paths. Coherent scattering and binding effects were considered in the simulations, which make the results suitable for calculations with the most complete descriptions of the attenuation coefficients. Differences due to geometry, source directionality, and detector response variability were typically found to be up to 30 \% in the studied range.
\end{abstract}


\end{frontmatter}

\section{Introduction}
The need of accounting for scattered radiation contribution to dosimetric magnitudes lead to the introduction of the so called ``build-up factor'' \cite{white:1950,harima:1993}. Multiplication by this magnitude, which is regarded as a function of the source energy and the medium thickness, allows correcting the exponential propagation model, which takes only into account the contribution of uncollided source particles.

Traditional calculations of the build-up factor typically involved defining a numerical method to address a simplified transport problem where some phenomena (like ionized electron bremsstrahlung or coherent scattering) are left out for calculations to be feasible. Among those, the method of moments \cite{goldstein:1954}, the discrete ordinate-integral transport \cite{takeuchi:1984}, the anisotropic source-flux iteration technique \cite{gopinath:1971}, and the method of invariant embedding \cite{shimizu:2002} should be highlighted. A general review of these techniques can be found in \refcite{harima:1993}. Nowadays, the most extended method to address particle transport avoiding simplifying the physics is the Monte Carlo simulation. As a consequence, most of the recent developments resort to this technique to calculate the build-up factors \cite{chibani:2001,ouahab:2011,namito:2012,sardari:2009}.

The main weakness of the build-up formalism is the multiple dependences that are hidden beneath it, the most relevant among them being \cite{harima:1993} the dosimetric magnitude studied (air-kerma, dose in tissue or other media...), angular and spatial distribution of the source, and geometry of the medium. Application of an inadequate definition of the build-up factor might thus lead to wrong values in the resulting calculations, as shown later in this work.

The purpose of this work is to present a general formulation to address the dependence on the response characterizing the dosimetric magnitudes, which will lead to the definition of the differential build-up factor. A similar notion can be found in \refcite{shimizu:2002}, but here we will show a more rigorous definition which also covers general sources and geometries. This magnitude will be evaluated for different materials, with energies ranging from 30 keV to 10 MeV, in depths up to 10 mean-free-paths using analog Monte Carlo techniques, including an analysis of the error arising from them. The calculations will include the effects of coherent scattering and binding effects in Compton scattering, so the build-up factors derived from it can be used with a full description of the attenuation coefficients, where those effects are included \cite{hubbell:1996}.

\section{Formalism}
The fundamental magnitude to describe a stationary linear transport problem is the angular flux $\psi(\bm r, E, \bm \Omega)$, where $\bm r$ is a point of space, $\bm \Omega$ is a solid angle, and $E$ is an energy.

Once the transport problem is solved, angular dependence becomes irrelevant and one may just consider the differential fluence in a point, which is given by
\begin{equation}\label{eq:angular-flux}
 \phi_\text{en}(\bm r, E)=\int_{\mathbb{S}^2} \psi(\bm r, E, \bm \Omega) \dif \bm \Omega
 \,.
\end{equation}

Dosimetric magnitudes are found by weighted integrals of the differential fluence, i.e.,
\begin{equation}\label{eq:dosimetric-magnitude}
 D_i(\bm r) = \int_{\mathbb{R}^+} \phi_\text{en}(\bm r,E) c_i(E) \dif E
 \,,
\end{equation}
where $c_i$ is a function which defines each dosimetric magnitude $D_i$. Some common cases are shown in~\tabref{tab:functionals}, where $\mu_\text{en}$ is the energy absorption coefficient \cite{seltzer:1993,hubbell:1996}. Other dosimetric magnitudes, like the air kerma \cite{johns:1969} or the ambient dose equivalent \cite{ferrari:1994} can be obtained with their specific conversion coefficients. 

\begin{table}[h!]
	\centering
		\begin{tabular} {cc}
		Functional & $c_i(E)$ \\
		\hline
		(Total) (number) fluence & 1  \\
		(Total) energy fluence & E \\
		Dose & $E \cdot \mu_{\text{en}}(E) \rho^{-1}$ \\
		\end{tabular}
	\caption{Some dosimetric magnitudes that can be defined using~\eqref{eq:dosimetric-magnitude}. Words in brackets are sometimes omitted.}
	\label{tab:functionals}
\end{table}

The simplest model that can be used to estimate the value of a dosimetric magnitude after crossing a material is given by the so-called exponential model, where all scattered radiation is assumed not to contribute to the magnitude. In such a case, a monoenergetic source of energy $E_0$ behaves like
\begin{equation}\label{eq:exp-model}
D_i^0(\bm{r}) =\phi^0(\bm r) c_i(E_0)\,,
\end{equation}
where the superindex $0$ refers to the value of a magnitude due to collisionless contributions. This collisionless fluence can be calculated considering each source element decays exponentially with depth multiplied by an energy-dependent attenuation coefficient \cite{hubbell:1996}, which will be denoted as $\mu$.

Three different geometries will be considered. First, for a monodirectional source and a planar shielding \eqref{eq:exp-model} becomes
\begin{equation}
 D_i^0(\bm{r}) =\phi(\bm0) c_i(E_0) \eu^{-\mu(E_0) x}
=D_i^0(\bm0) \eu^{-\mu(E_0) x}      
\,,
\end{equation}
where $x$ is the coordinate associated to the depth of the shielding.
Second, an isotropic point source with an spherical shielding \cite{lamarsh:2001}, where \eqref{eq:exp-model} takes the form
\begin{equation}
 D_i^0(\bm{r})  = \frac{\phi(R)}{(r/R)^2} c_i(E_0) \eu^{-\mu(E_0) r}
	      = \frac{D_i^0(R) R^2}{r^2} \eu^{-\mu(E_0) r}      
\,,
\end{equation}
where $r$ is the radius of the spherical shielding, $R$ is an arbitrary distance used for normalization and $r, R>0$. Third, a planar distribution of isotropic sources crossing a planar shielding, where \eqref{eq:exp-model} turns out to be
\begin{equation}
D_i^0(\bm{r})=\frac{D_i^0(X)}{\expone{(\mu(E_0) X)}} \expone{(\mu(E_0) x)}    
\,,
\end{equation}
where $\expone$ is the principal value of the exponential integral as defined in \refcite{NIST:DLMF}, $x$ is the coordinate associated to the depth of the shielding, X is an arbitrary depth used for normalization, and $x, X>0$.

The build-up factor is usually introduced simply as a factor that corrects \eqref{eq:exp-model} to take into account the scattered radiation reaching the detector \cite{martin:2006,lamarsh:2001}. For a better understanding, we may recall first that the transport problem is described by the Boltzmann transport equation \cite{harima:1993}, which describes the distribution of the angular flux, and hence of the different dosimetric magnitudes by using \eqref{eq:angular-flux} and \eqref{eq:dosimetric-magnitude}.  The solution of this equation can also be stated as a transformation of the uncollided angular flux to the actual angular flux, from where dosimetric magnitudes are obtained again by integration. This can be put in the diagram
\begin{equation}
\begin{tikzcd}
\psi^0(\bm{r},E,\bm{\Omega}) \arrow[r, "\mathfrak{B}"] \arrow[d,"\iint c_i(E)"]
& \psi(\bm{r},E,\bm{\Omega}) \arrow[d, "\iint c_i(E)"] \\
D_i^0(\bm{r}) \arrow[r, "\mathcal{B}_i" ]
& D_i(\bm{r})
\end{tikzcd}
\,,
\end{equation}
where $\mathfrak{B}$ is the build-up operator, which is a Neumann series of an integro-differential operator \cite{jha:12}; and $\mathcal{B}_i$ is the build-up factor of the dosimetric magnitude $i$. Monte Carlo transport effectively approximates $\mathfrak{B}$ by sampling methods and any $\mathcal{B}_i$ can as well be obtained by integrating the results as shown in the diagram.

The restriction of the operator $\mathcal{B}_i$ to a given point $\bm r$ where the dosimetric magnitude is not zero can be identified with a product
\begin{equation}\label{eq:build_up}
 D_i(\bm{r})=B_i(\bm r, E_0) D_i^0(\bm{r}) 
 \,,
\end{equation}
where the dependence on the source energy (which is assumed to be monoenergetic) has been explicitly stated. Other dependencies such as source directionality or geometrical configuration are implicitly assumed though.  These results are usually tabulated, thus obtaining the typical build-up description.

In this work we suggest extending the build-up description by keeping the energy information, i.e., by not integrating in the energies, as shown in the diagram
\begin{equation}
\begin{tikzcd}
\psi^0(\bm{r},E,\bm{\Omega}) \arrow[r, "\mathfrak{B}"] \arrow[d,"\int"]
& \psi(\bm{r},E,\bm{\Omega}) \arrow[d, "\int"] \\
\phi_\text{en}^0(\bm{r};E) \arrow[r, "\mathcal{B}" ]
& \phi_\text{en}(\bm{r};E)
\end{tikzcd}
\,,
\end{equation}
where $\mathcal{B}$, what we shall call the differential build-up operator, is defined by the diagram.

Its form is more complicated than a simple energy-dependent factor. Considering linear superposition on the source energy must hold, we have
\begin{equation}\label{eq:diff_build_up}
 \phi_\text{en}(\bm{r};E)=\int_0^\infty \mathcal{B}(\bm r, E'\rightarrow E) \phi_\text{en}^0(\bm{r};E') \dif E' 
 \,,
\end{equation}
where $\mathcal{B}(\bm r, E'\rightarrow E) \dif E \dif E'$ is the amount of particles with energies infinitesimally close to $E$ which arrive per uncollided particle arriving with an energy infinitesimally close to $E'$ (which was also the source emission energy). Hence, $\mathcal{B}(\bm r, E'\rightarrow \variable)$ is a generalized function. The application of \eqref{eq:diff_build_up} to a monoenergetic source immediately describes $\mathcal{B}(\bm r, E'\rightarrow \variable)$ as the quotient of the differential fluence produced by such a source, divided by the total uncollided fluence. These distributions can be approximated by histograms or other similar techniques to provide a practical description of the operator.

From a more mathematical point of view, the operator $\mathcal{B}$ in \eqref{eq:diff_build_up} is actually mapping probability distributions to probability distributions (as well as changing an additional normalization), and hence we are assuming here that the $\mathcal{B}(\bm r, E'\rightarrow \variable)$ might have a discrete component which we admit to be described by one or more Dirac delta distributions added to the continuous component. More formal ---but equivalent in practice--- approaches are those usually taken in probability theory, like splitting \eqref{eq:diff_build_up} into a ``density'' integral and ``mass'' sum or defining a Stieltjes integral instead \cite{feller:2008}.

Finally, the usual build-up factors can be recovered by taking the product of \eqref{eq:diff_build_up} with $c_i(E)$, using a monoenergetic source, and integrating in the source energy $E'$, i.e.,
\begin{equation}
\begin{split}
 D_i(\bm r) &= \int_0^\infty \int_0^\infty \mathcal{B}(\bm r, E'\rightarrow E) \phi_\text{en}^0(\bm{r};E') c_i(E) \dif E \dif E' \\
	    &= \int_0^\infty \mathcal{B}(\bm r, E_0\rightarrow E) \phi^0(\bm{r}) c_i(E) \frac{c_i(E_0)}{c_i(E_0)} \dif E \\
	    &= D_i^0(\bm r) \frac{\int_0^\infty \mathcal{B}(\bm r, E_0\rightarrow E) c_i(E )\dif E}{c_i(E_0)}
 \,.
\end{split}
\end{equation}
By comparison with \eqref{eq:build_up} one finds
\begin{equation}\label{eq:recover_build_up}
B_i(\bm r,E_0)= \frac{\int_0^\infty \mathcal{B}(\bm r, E_0\rightarrow E) c_i(E )\dif E}{c_i(E_0)}
\,,
\end{equation}
so the traditional build-factors are just weighted integrals of the differential build-up factor.

Both the differential build-up factor and the usual build-up factors can be extended to describe energy-distributed sources. The mathematical details are given in the appendix.

\section{Simulation description}\label{sec:simulation}
\begin{figure}
\includegraphics[width=.98\linewidth]{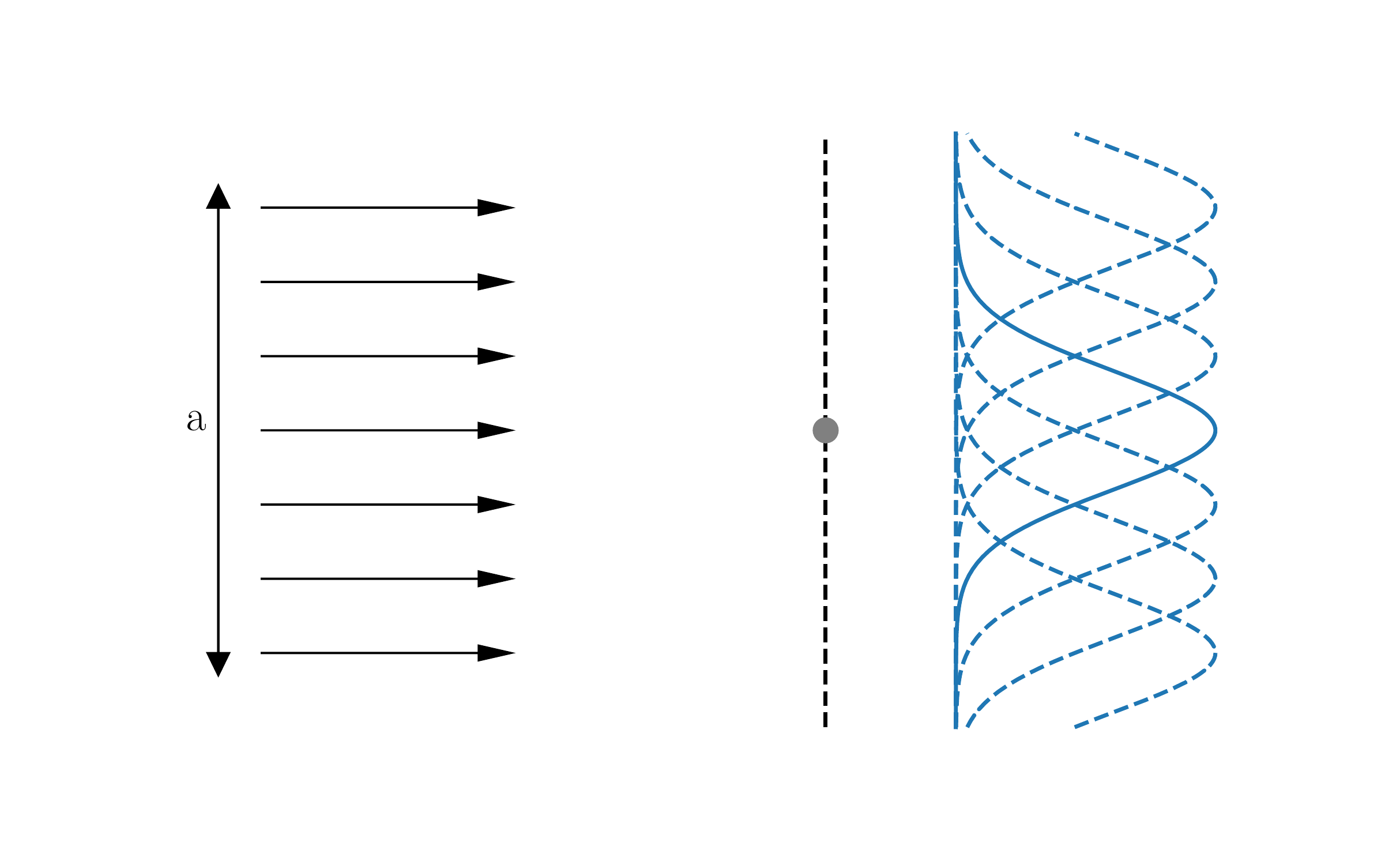}
\caption{\label{fig:geometry-a}Schematic view of the geometry used to define the build-up factor in a planar geometry. Contributions from a source with size $a$ are scored in much smaller detector, here regarded as punctual. Contributions from each element of the source are shown in blue.}
\end{figure}

The two planar geometries described before (with monodirectional and isotropic sources) are defined with an extended source whose effects on a much smaller detector are considered. This situation is depicted in \figref{fig:geometry-a}.

\begin{figure}
\includegraphics[width=.98\linewidth]{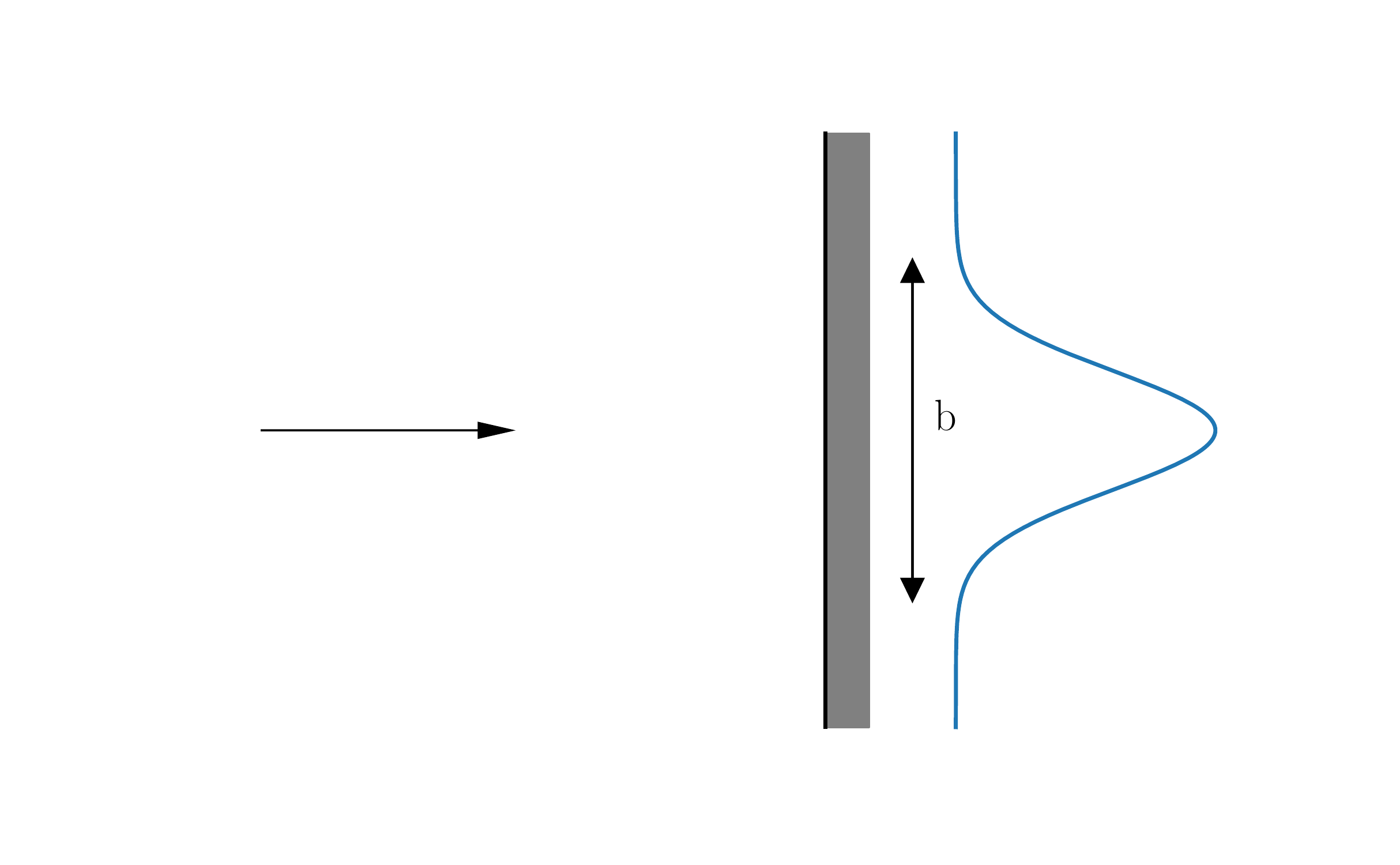}
\caption{\label{fig:geometry-b}Schematic view of the geometry used to calculate the build-up factor in a planar geometry. Contributions from a punctual source are scored in a planar detector. If the extent of its effect can be described by a length $b$, this configuration might represent the one in \figref{fig:geometry-a} as long as $a>b$.}
\end{figure}

A more practical calculation can be done considering a punctual source, as shown in \figref{fig:geometry-b}, and measuring the build-up in the whole plane. The extended source is a superposition of this punctual kernel, also shown in \figref{fig:geometry-b}. In the monodirectional case, a simple change of variable shows both results are related by the density of the source, and hence the planar integration of the punctual source is dual to the punctual measuring of the planar source, making the computations more efficient. The isotropic case is not complicated either, calculations done with detail can be found in \refcite{namito:2012}.

The punctual isotropic source with spherical shielding can be simulated with no further transformation. It is somewhere stated \cite{lamarsh:2001} that its results can be used by superposition to represent other geometries like the planar-isotropic. However, this can only be understood as an approximation since the geometry seen by each component of the source is different from this idealization.

All the simulations were done with the Monte Carlo package FLUKA 2011.2x.3 \cite{battistoni:2007, bohlen:2014}. Each of the simulations consisted of between $3\cdot{10}^6$ and $6\cdot{10}^6$ primary particles, propagating in the described geometries of aluminum, iron, lead, water, air, and concrete. 

The differential fluence across the surfaces was measured using a plane-crossing detector (\text{USRDBX} in the FLUKA terminology). A previously tested user routine \cite{hernandez:2016} was used to discard contributions in each of the surfaces coming from particles which come from stories that originated beyond that surface, thus effectively simulating the transmission problem in all the depths with a single simulation. These detectors were placed up to a depth of 10 mean-free-paths (mfp) ---the inverse of the attenuation coefficient--- in steps of 0.1 mfp.

The simulated physics included Rayleigh scattering, form factor-corrected Compton
scattering, and fluorescence emission. All transport and production thresholds were set at \SI{1}{keV} for both photons and electrons.

There are two sources of error worth considering in the estimation of $\mathcal{B}$ and the $\mathcal{B}_i$ derived from it. On the one hand, there is a discretization error due to approximating the energy distribution with a mesh. Thus, the exact expression of \eqref{eq:recover_build_up} is replaced by a finite sum
\begin{equation}\label{eq:recover_build_up_discrete}
B_i(\bm r,E_0) \approx \frac{\sum_j c_i(\tilde{E}_j) \int_{E_j}^{E_{j+1}}\mathcal{B}(\bm r, E_0\rightarrow E) \dif E}{c_i(E_0)}
\,,
\end{equation}
where the integral is the object being estimated by the Monte Carlo method and $\tilde{E}_j$ is an energy value chosen to represent the interval $\left[E_j, E_{j+1}\right]$. The natural choice is the mean value of the interval, except for the bin with the highest energies, where most of the particles will have experienced no collisions, so in this case the source energy should be used instead. The error due to this discretization is bounded by application of \eqref{eq:recover_build_up_discrete} itself, choosing the maximum and minimum of $c_i$ in each interval.

On the other hand, there is a statistical error inherent to the Monte Carlo method, which is described by the error of the mean of the integral being sampled. Assuming this distribution is normal, \eqref{eq:recover_build_up_discrete} is also describing a weighted sum of Gaussians. The resulting statistical error of a magnitude $i$ is thus obtained with the well known expression
\begin{equation}
\sigma=\sqrt{\sum_j \left(c_i(\tilde{E}_j) \delta_j \right)^2}
\,,
\end{equation}
where $\delta_j$ is the estimated typical deviation of the mean number of counts in the interval $\left[E_j, E_{j+1}\right]$.

In the following, all error bars will represent $2\sigma$ plus the discretization error. When a quotient of build-up factors is represented, the statistical error of the ratio will be derived from the Taylor expansion of the quotient of Gaussians \cite{taylor:1996}, while the discretization error will be obtained from interval arithmetic \cite{hickey:2001}.

\section{Results}
\begin{figure}
\includegraphics[width=.98\linewidth]{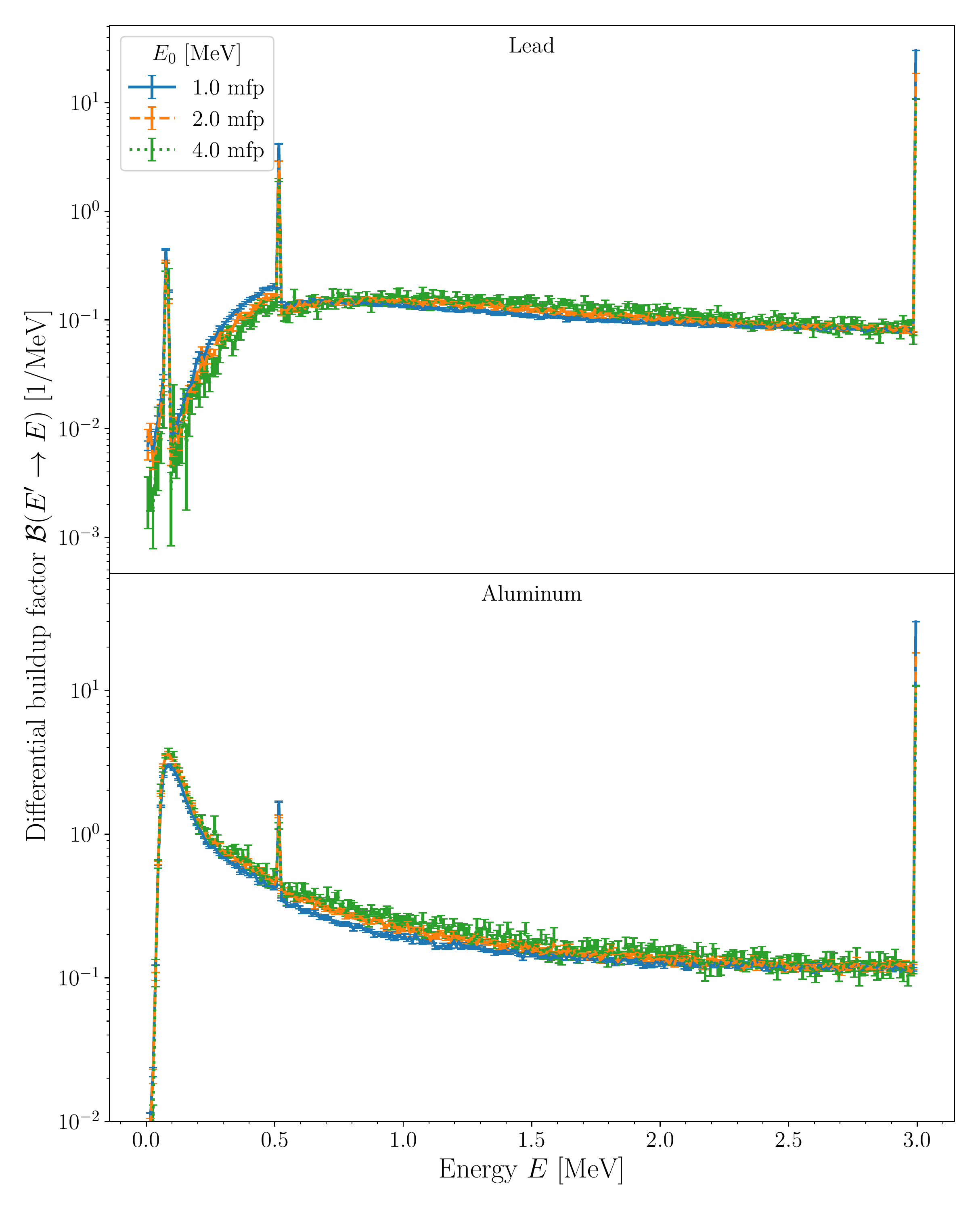}
\caption{\label{fig:diff-buildup}Differential buildup factor as a function of the arriving photon energy $E$ for the planar-isotropic geometry in lead (above) and aluminum (below), with a \SI{3}{MeV} source. Each of the series represent a certain thickness, as described in the legend. The peaks can be identified from left to right with the K-edge of lead ($\sim\SI{88}{keV}$, only above), the annihilation energy ($\sim\SI{511}{keV}$) and  the source energy.}
\end{figure}

The differential build-up factor obtained from a planar-isotropic source is shown in \figref{fig:diff-buildup}. The error bars in the figure show the statistical error estimated by the Monte Carlo sampling. The curves shown there are proportional to the measured differential fluence. The source energy, the annihilation energy ($\sim\SI{511}{keV}$) and the K-edge of lead ($\sim\SI{88}{keV}$) can be identified in the figure. It is worth noting that the relative height of the bins containing these energies is  highly dependent on the mesh, due to the discrete nature of this contributions. The general results are available online \cite{repo}.

\begin{figure}
\includegraphics[width=.98\linewidth]{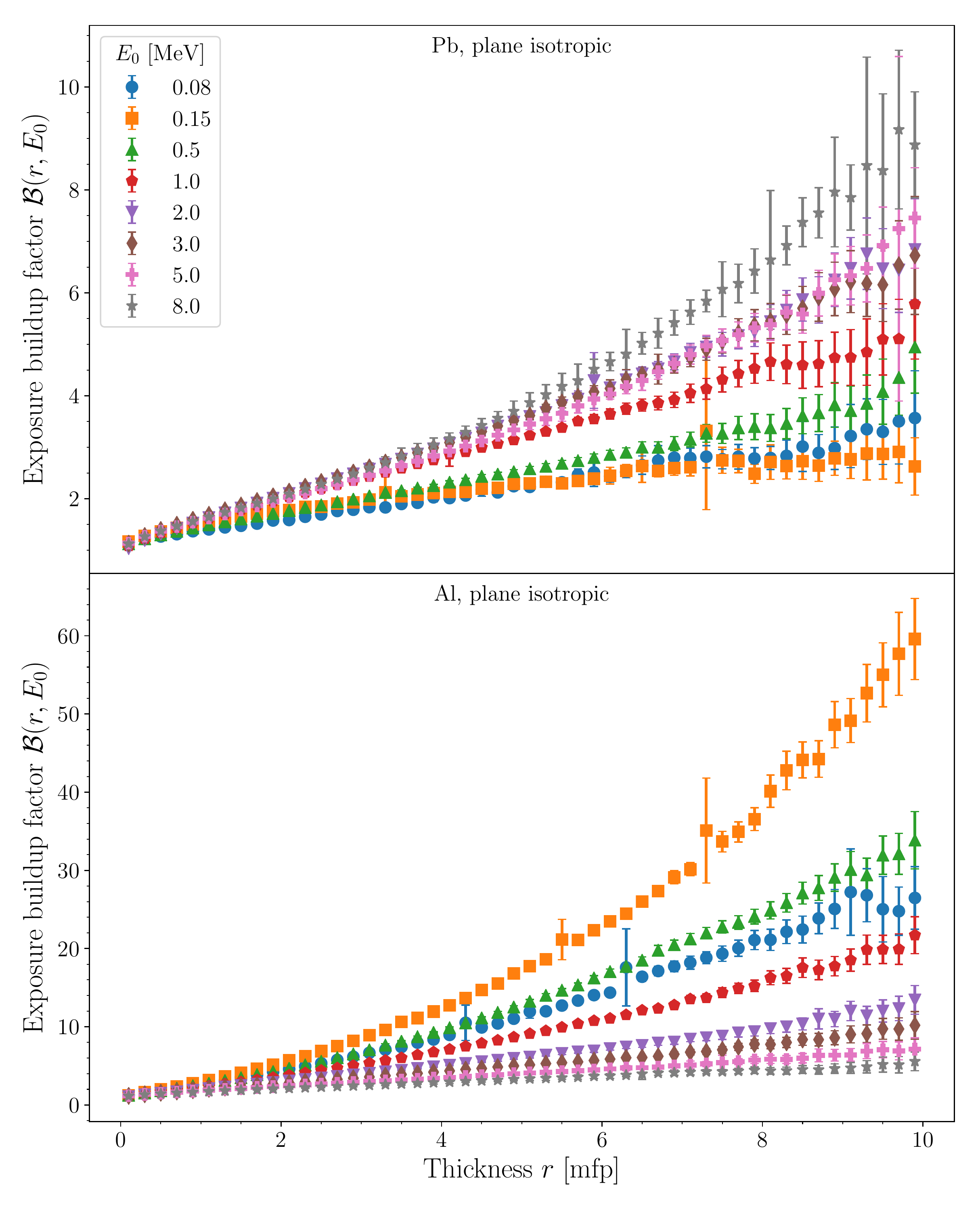}
\caption{\label{fig:exposure-buildup}Exposure build-up factor as a function of the thickness of the material, for the planar-isotropic geometry in lead (above) and aluminum (below). Each of the series represent a certain source energy, as described in the legend.}
\end{figure}

The usual build-up factors can be obtained by integration of \eqref{eq:recover_build_up}. If a coarser energy grid is used, it would be important for the bin with the source energy to accurately represent it, as discussed before. Some of the series in the source energy of the exposure build-up factor are shown in \figref{fig:exposure-buildup}, also for lead and aluminum. Some of the results are tabulated in \tabref{tab:exposure-planariso} for informative purpose on the source of the errors.

\begin{table*}\tiny
\begin{tabular}{cccccccc}
& \multicolumn{7}{c}{Energies [MeV]}\\
$\mu x$ & 1 & 2 & 3 & 4 & 6 & 8 & 10\\
 0.08 & $1.08\pm0.00\pm0.00$ & $1.03\pm0.00\pm0.00$ & $1.13\pm0.00\pm0.01$ & $1.09\pm0.00\pm0.02$ & $1.11\pm0.00\pm0.03$ & $1.12\pm0.00\pm0.04$ & $1.07\pm0.00\pm0.05$\\
 0.5 & $1.34\pm0.00\pm0.00$ & $1.31\pm0.00\pm0.00$ & $1.39\pm0.00\pm0.01$ & $1.34\pm0.00\pm0.03$ & $1.35\pm0.00\pm0.04$ & $1.37\pm0.00\pm0.06$ & $1.35\pm0.00\pm0.07$\\
 1.0 & $1.58\pm0.00\pm0.00$ & $1.58\pm0.00\pm0.00$ & $1.64\pm0.00\pm0.01$ & $1.57\pm0.00\pm0.04$ & $1.56\pm0.00\pm0.04$ & $1.60\pm0.00\pm0.06$ & $1.62\pm0.00\pm0.09$\\
 2.0 & $1.99\pm0.00\pm0.00$ & $2.06\pm0.00\pm0.01$ & $2.09\pm0.00\pm0.02$ & $1.99\pm0.00\pm0.04$ & $1.96\pm0.00\pm0.07$ & $2.04\pm0.00\pm0.07$ & $2.15\pm0.00\pm0.11$\\
 3.0 & $2.40\pm0.01\pm0.03$ & $2.54\pm0.01\pm0.00$ & $2.56\pm0.01\pm0.02$ & $2.43\pm0.01\pm0.06$ & $2.38\pm0.01\pm0.08$ & $2.53\pm0.01\pm0.09$ & $2.77\pm0.01\pm0.14$\\
 5.0 & $3.18\pm0.04\pm0.00$ & $3.52\pm0.03\pm0.01$ & $3.57\pm0.03\pm0.06$ & $3.43\pm0.04\pm0.21$ & $3.44\pm0.03\pm0.04$ & $3.74\pm0.03\pm0.18$ & $4.42\pm0.03\pm0.15$\\
 8.0 & $4.72\pm0.24\pm0.01$ & $5.39\pm0.20\pm0.01$ & $5.57\pm0.23\pm0.01$ & $5.27\pm0.19\pm0.01$ & $5.24\pm0.20\pm0.02$ & $6.46\pm0.21\pm0.04$ & $8.68\pm0.21\pm0.06$\\
 10.0 & $5.51\pm0.66\pm0.01$ & $7.61\pm0.77\pm0.01$ & $7.09\pm0.72\pm0.02$ & $6.07\pm0.65\pm0.02$ & $7.73\pm0.64\pm0.03$ & $8.33\pm0.61\pm0.05$ & $13.89\pm0.71\pm0.10$\\
  \end{tabular}
 \caption{\label{tab:exposure-planariso}Build-up factor for the exposure of a planar-isotropic source in lead. The first number indicates the value of the magnitude, the second one standard deviation of the Monte Carlo uncertainty, and the third one the error bound due to discretization.}
\end{table*}

\begin{figure}
\includegraphics[width=.98\linewidth]{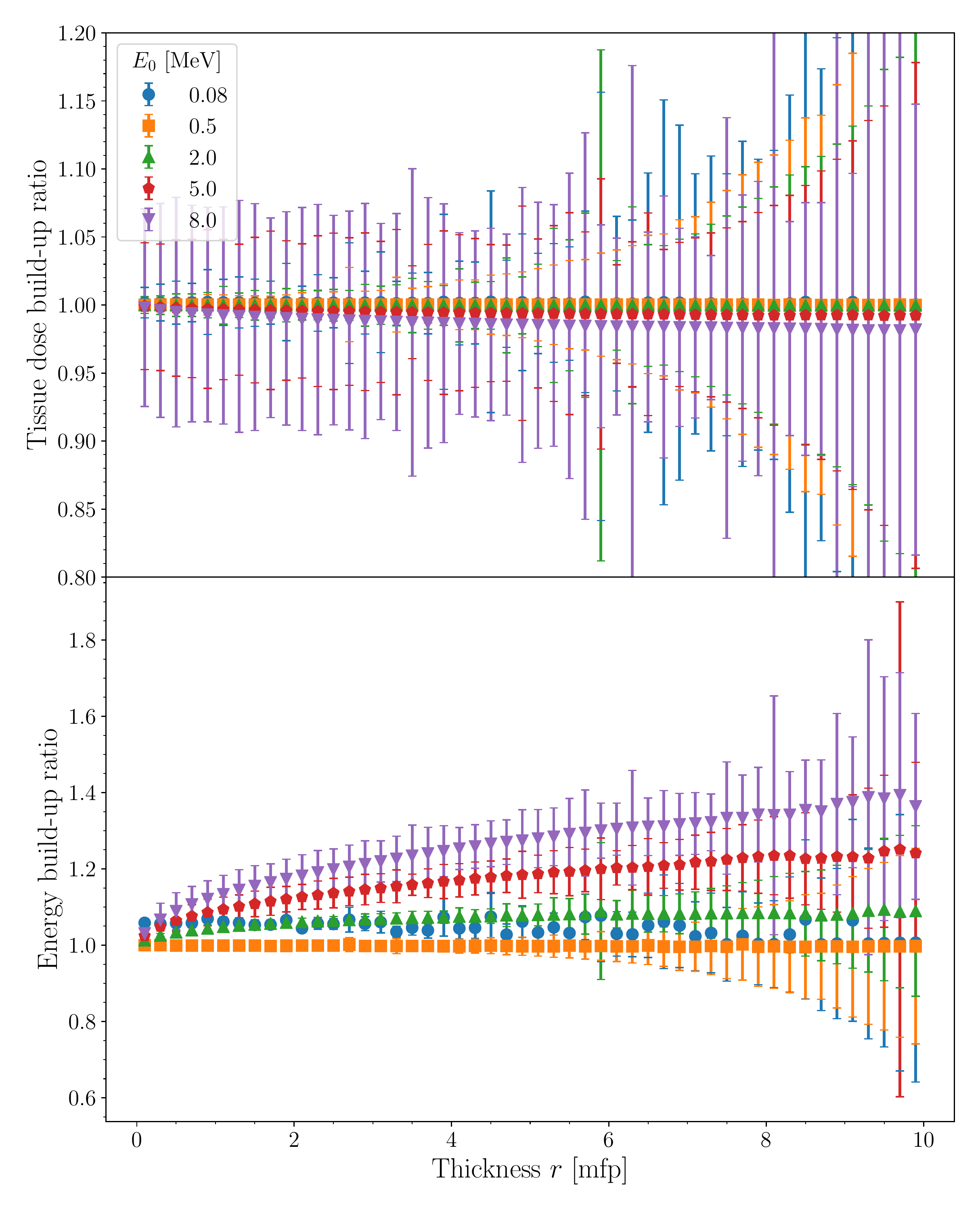}
\caption{\label{fig:differences-magnitude}Ratio of the build-up factors of dose in tissue (above) and energy fluence (below) with the air-exposure build-up factor, for the planar-isotropic geometry in lead as a function of the thickness of the material. Each of the series represent a certain source energy, as described in the legend.}
\end{figure}

The relative change of the build-up factor with the dosimetric magnitude is shown in \figref{fig:differences-magnitude}, where lead was chosen as the material. The difference between exposure and dose in tissue is below the error bars, which are around 10\%. The difference between energy and exposure build-ups is more noticeable, being up to around 30\%. The plots from the other materials or with different energy series would be similar to \figref{fig:differences-magnitude}, thus leading to the same conclusions.

\begin{figure}
\includegraphics[width=.98\linewidth]{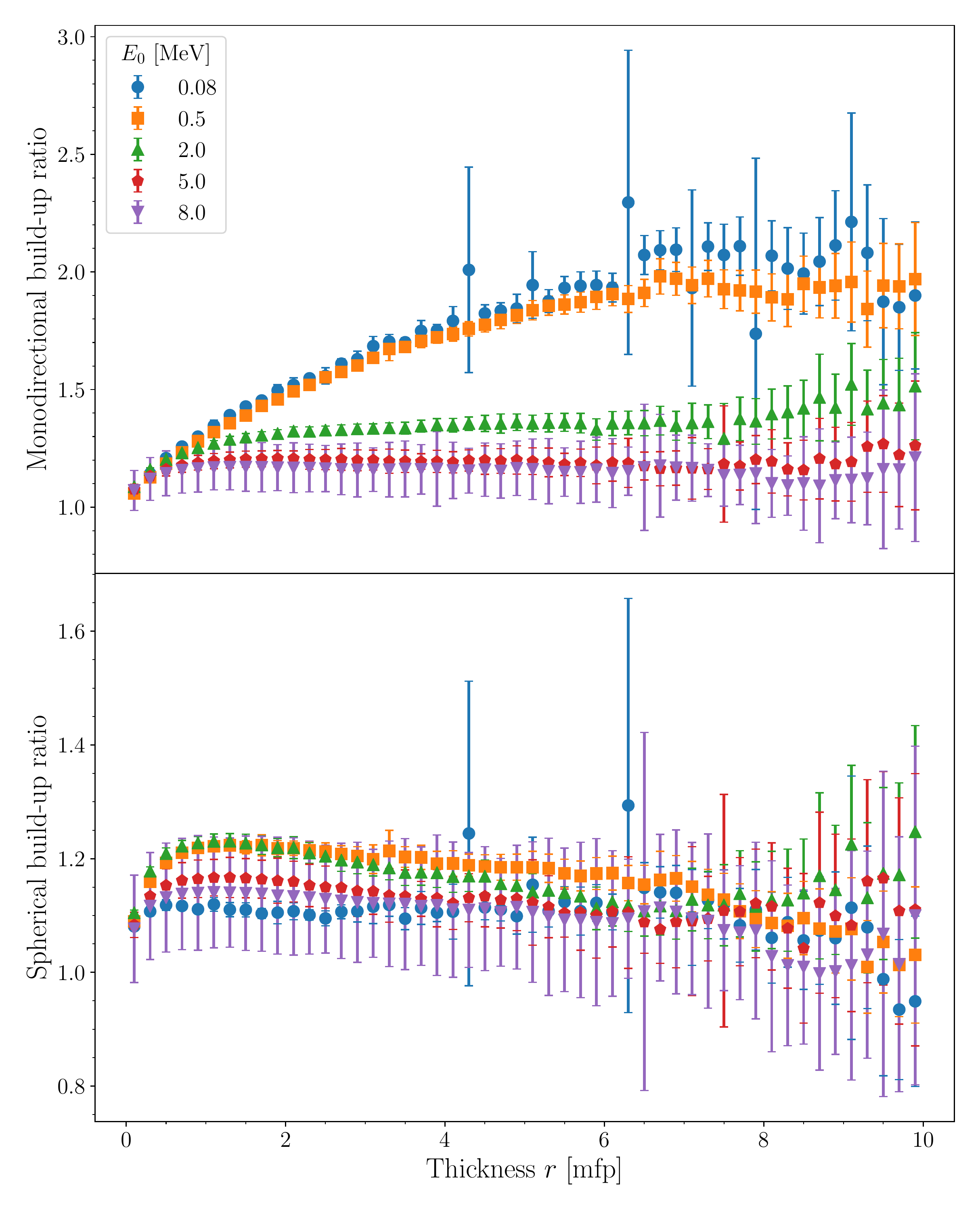}
\caption{\label{fig:differences-geometry}Ratio of the exposure build-up factors of the planar-monodirectional (above) and spherical-isotropic (below) geometries with the planar-isotropic geometry in lead, as a function of the thickness of the material. Each of the series represent a certain source energy, as described in the legend.}
\end{figure}

Analogously, the change due to the geometric considerations is shown in \figref{fig:differences-geometry}, again for lead. Differences between the planar-isotropic and the planar-monodirectional geometries can be of about 200 \%, while differences between the planar-isotropic and the spherical-isotropic geometries are up to around 30 \%.  The plots for the other materials or with different energy series would be similar to \figref{fig:differences-magnitude}, although in some cases the differences with the planar-monodirectional geometry might increase up to around 400 \%.

\begin{figure}
\includegraphics[width=.98\linewidth]{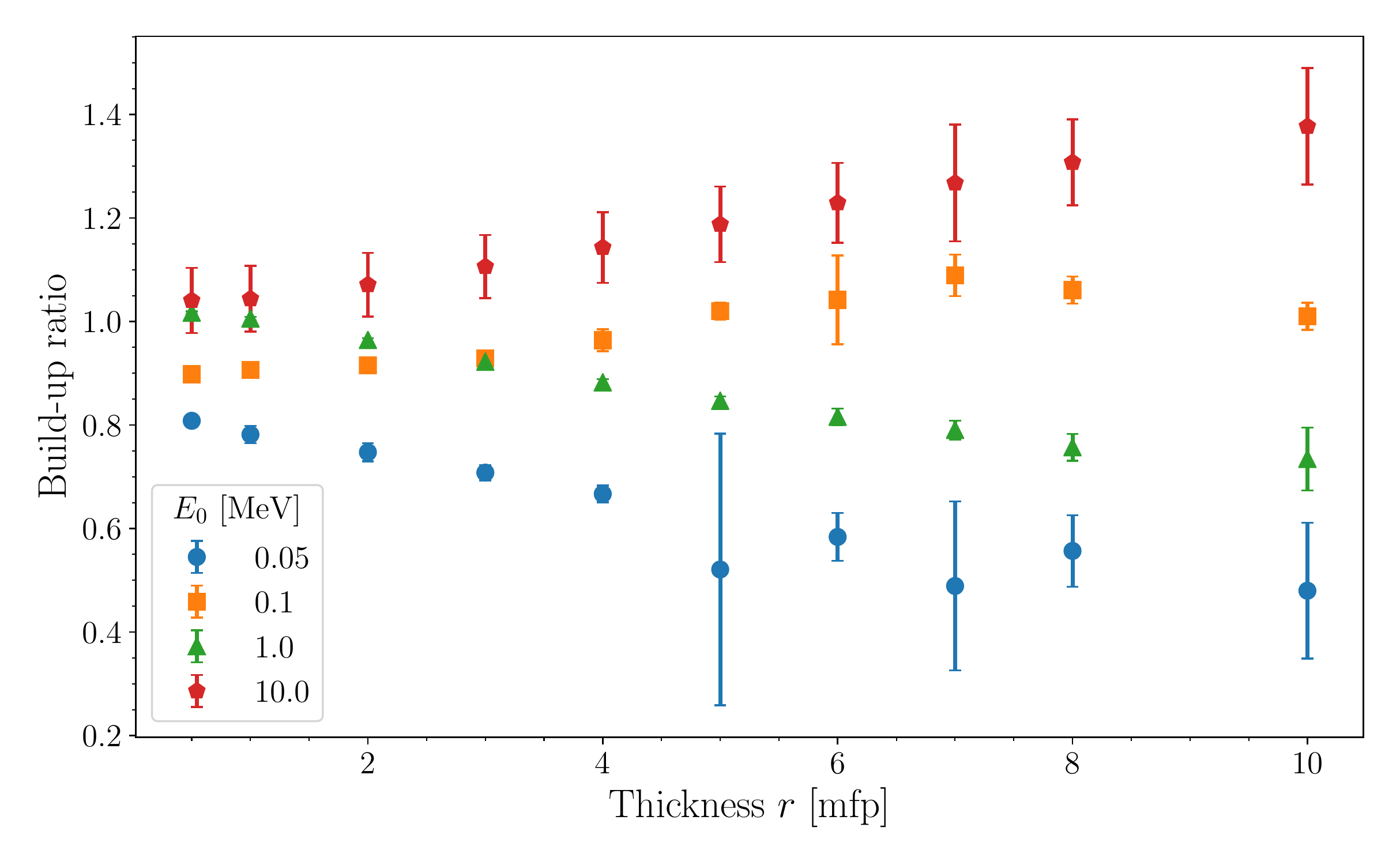}
\caption{\label{fig:ratio-ans}Ratio of the exposure build-up factors of lead from \protect\refcite{harima:1993} with the ones derived in the current work for the spherical-isotropic geometry. Each of the series represent a certain source energy, as described in the legend.}
\end{figure}

Differences with older build-up factor calculations due to bremsstrahlung, Rayleigh scattering, form factor-corrected Compton
scattering, and fluorescence emission is a subject that has already been studied in the literature, and shown relevant for different materials and energy ranges \cite{harima:1993, chibani:2001}. As an example of these differences, the ratios of the ANS/ANS-6.4.3 data, obtained from \refcite{harima:1993}, with the analogous spherical-isotropic results in lead is depicted in \figref{fig:ratio-ans}. The magnitude and behavior of this ratio are similar to those appearing in \refcite{chibani:2001}.

\section{Conclusions}
The differential build-up factor, a mathematical object able to describe the build-up regardless of the response function, has been introduced. Traditional build-up factors can be recovered as its weighted integrals. Analog Monte Carlo simulations were used to calculate the build-up up to 10 mean-free-paths for source energies in the 30 keV to 10 MeV range, using the most commonly studied materials, considering different hypothesis on the geometry and the source. Error bounds are also available for these calculations. Differences due to geometry and detector response were found to be typically up to 30 \% in the studied range, while differences with the source directionality might up to 200 \% or 400\% depending on the material. Coherent scattering and binding effects in Compton scattering were accounted for in the simulations, which makes the derived tabulations suitable for attenuation coefficients where these effects are also included. The results for aluminum, iron, lead, water, air, and concrete are available online \cite{repo}. This repository will be extended in the future to cover more materials and depths.

\section*{Acknowledgments}
One of the authors (G. H.) gratefully acknowledges the Consejer\'ia de Educaci\'on de la Junta de Castilla y Le\'on and the European Social Fund for financial support.

\appendix
\section{Non-monoenergetic sources}
Despite the fact that the build-up factor is usually applied to monoenergetic sources, it has been pointed \cite{harima:1993} that this can be regarded as an additional dependence in the build-up factor definition which can be considered. Furthermore, some specific calculations for continuous sources can be found in the literature \cite{rafiei:2018}. For the sake of completeness, we present here how the build-up factors for a general energy distribution of the source can be derived from the monoenergetic ones, using a planar-monodirectional geometry as an example.

The exponential model for a monodirectional source in a planar geometry with an energy distribution $S$ is given by linear superposition of \eqref{eq:exp-model}, \ie
\begin{equation}\label{eq:exp-model-distribution}
D_i^0(\bm{r}) = \int_0^\infty \phi(\bm0) S (E_0) c_i(E_0) \eu^{-\mu(E_0) r} \dif E_0  \,.
\end{equation}

The build-up factor can be defined taking the quotient of \eqref{eq:exp-model-distribution} with an analogous linear superposition that includes the monoenergetic build-up. In order to obtain a clear expression, we shall introduce the function
\begin{equation}\label{eq:weight-distribution}
w_i^S\left(\bm{r}, E_0\right) \defeq S(E_0) \eu^{-\mu(E_0)r}c_i(E_0)
\end{equation}
which considers the changes of the weights due to the change of the spectrum in the exponential model with depth, as well as the attenuation itself. Using this quantity, we can obtain the distribution build-up factors as
\begin{equation}\label{eq:buildup-distribution}
B_i\left[\bm{r}, S\right]= \frac{\int_0^\infty B_i(\bm r, E_0) w_i^S\left(\bm{r}, E_0\right) \dif E_0}{\int_0^\infty w_i^S\left(\bm{r}, E_0\right) \dif E_0}\,.
\end{equation}

Following the same reasoning it can be shown that \eqref{eq:buildup-distribution} is also valid for the point isotropic source with spherical shielding. However, in the case of a planar-isotropic source the exponential in \eqref{eq:weight-distribution} must be replaced with a $\expone$ function.

The differential build-up factor could also be defined for distributions, but it is just a linear superposition with $S$, as discussed previously.

\bibliography{diffbuildup}

\end{document}